\documentclass[aps,pra,showpacs,twocolumn]{revtex4-1}
\usepackage{amssymb}
\usepackage{amsmath}
\usepackage{graphicx}
\usepackage{epsfig}
\usepackage{subfigure}
\usepackage{amsfonts}
\usepackage{color}

\pdfoutput=1
\usepackage{float}
\usepackage{xcolor}
\usepackage{graphicx}
\usepackage{dcolumn}
\usepackage{bm,lipsum}
\usepackage[colorlinks=true, linkcolor=blue, citecolor=blue, urlcolor=blue]{hyperref}

\usepackage{graphicx}
\usepackage{booktabs}
\usepackage{tabularx}
\usepackage{xcolor,colortbl}

\usepackage{pdfpages}
\makeatletter
\AtBeginDocument{\let\LS@rot\@undefined}
\makeatother

\definecolor{Gray}{gray}{0.9}

\newcolumntype{a}{>{\columncolor{Gray}}c}
\newcolumntype{b}{>{\columncolor{white}}c}

\begin{document}

\title{Quantum dynamics of dissipative Chern insulator}

\author{J.L. Zhong, X.Y. Li}
\affiliation{School of Physics and Electronic Engineering, Jiangsu University, Zhenjiang 212023, China}

\begin{abstract}
 For open quantum systems, a short-time evolution is usually well described by the effective non-Hermitian Hamiltonians, while long-time dynamics requires the Lindblad master equation,in which the Liouvillian superoperators characterize the time evolution. In this paper, we constructed an open system by adding suitable gain and loss operators to the Chen insulator to investigate the time evolution of quantum states at long times by numerical simulations. Finally, we also propose a topolectrical circuits to realize the dissipative system for experimental observation. It is found that the opening and closing of the Liouvillian gap leads to different damping behaviours of the system and that the presence of non-Hermitian skin effects leads to a phenomenon of chiral damping with sharp wavefronts. Our study deepens the understanding of  quantum dynamics of dissipative system.
\end{abstract}

\maketitle
\section{Introduction}
With the laboratory advances in modulating dissipation and quantum coherence,the theory of open and nonequilibrium systems has received renewed attention\cite{bergholtz2021exceptional,ashida2020non}.Non-Hermitian Hamiltonians have been used to describe a large number of non-conservative systems, such as classical waves with gain and loss\cite{zhu2014p,popa2014non,regensburger2012parity,feng2014single,zhou2018observation,cerjan2019experimental}, solids with finite quasiparticles lifetimes\cite{papaj2019nodal,shen2018quantum,cao2021non}, and open quantum systems\cite{rotter2009non,xiao2020non,xiao2021observation}.The unique features of non-Hermitian systems have been recognized in a variety of physical settings, in particular the non-Hermitian skin effect(NHSE)\cite{yao2018edge,yao2018non}, where the eigenstates of the system are exponentially localized on the boundary. In recent years, the impact of NHSE has been extensively studied\cite{lee2019anatomy,wang2019non,lee2019hybrid,kunst2019non,borgnia2020non,wang2018topological,ji2024non,li2022topological,ezawa2019braiding,yang2022non,ge2019topological,ji2024generalized,li2022universal}.

NHSE was also found in open quantum systems\cite{song2019non}. For open quantum systems, the non-Hermitian effective Hamiltonian describes the time evolution of the wavefunction under post-selection conditions, while the time evolution of the density matrix (without post-selection) is driven by the Liouvillian superoperator in the master equation\cite{ashida2020non,dalibard1992wave,carmichael1993quantum,weimer2021simulation}. It has been found that the Liouvillian superoperator can also exhibit non-Hermitian skin effects and that such effects can significantly affect the dynamical behaviour of the system at long times\cite{song2019non,cai2013algebraic,zhou2022non,li2022engineering,he2022damping,liu2020helical,luo2019higher,manzano2020short,haga2021liouvillian,mcdonald2020exponentially,mcdonald2022nonequilibrium}. In a large class of open quantum systems, the quantum state in the long time limit converges to the steady state by algebraic damping under periodic boundary conditions and exponential damping under open boundary conditions\cite{song2019non}.

In recent years, it has been discovered that topolectrical circuits can be used as platform to simulate the lattice systems, thus enabling the study of topological states in topolectrical circuits and gradually developing the field of topological circuitry\cite{lee2018topolectrical,imhof2018topolectrical,albert2015topological}. Some of the early experiments and theories were extensively studied in Hermitian systems\cite{imhof2018topolectrical,yang2020observation}. Since the phenomena of non-Hermitian systems are more rich than that of Hermitian systems, increasing attentions are contributed into the non-Hermitian physics, and some interesting phenomena have also been realized by topolectrical circuits\cite{helbig2020generalized,hofmann2019chiral,ezawa2019electric,ezawa2019non,schindler2012symmetric}.
\newline
Previous studies on open quantum dynamics and topolectrical circuits have mainly focused on one-dimensional non-Hermitian models, and relatively few studies on higher-dimensional non-Hermitian models. In this paper, we consider a two-dimensional open quantum system based on Chen insulators. Following the method developed in Ref.\cite{song2019non}, we study the dynamics of this system in terms of the damping matrix derived from the Liouvillian superoperator, and give a model of topolectrical circuit realization of the damping matrix based on Kirchhoff’s theory. It is found that due to the NHSE of the damping matrix, the long-time dynamics of the system under open boundary conditions is significantly different from that under periodic boundary conditions.

Our paper is organized as follows. In section II,we briefly review the general framework on how to convert Liouvillian operators with linear jumps to non-Hermitian damping matrix. In sections III and IV, we compute and numerically simulate the long-time evolution of the model. In section V we give the circuit model of the non-Hermitian damping matrix . Finally, we conclude in section VI.

\begin{figure*}[]
	\includegraphics[width=0.98\textwidth]{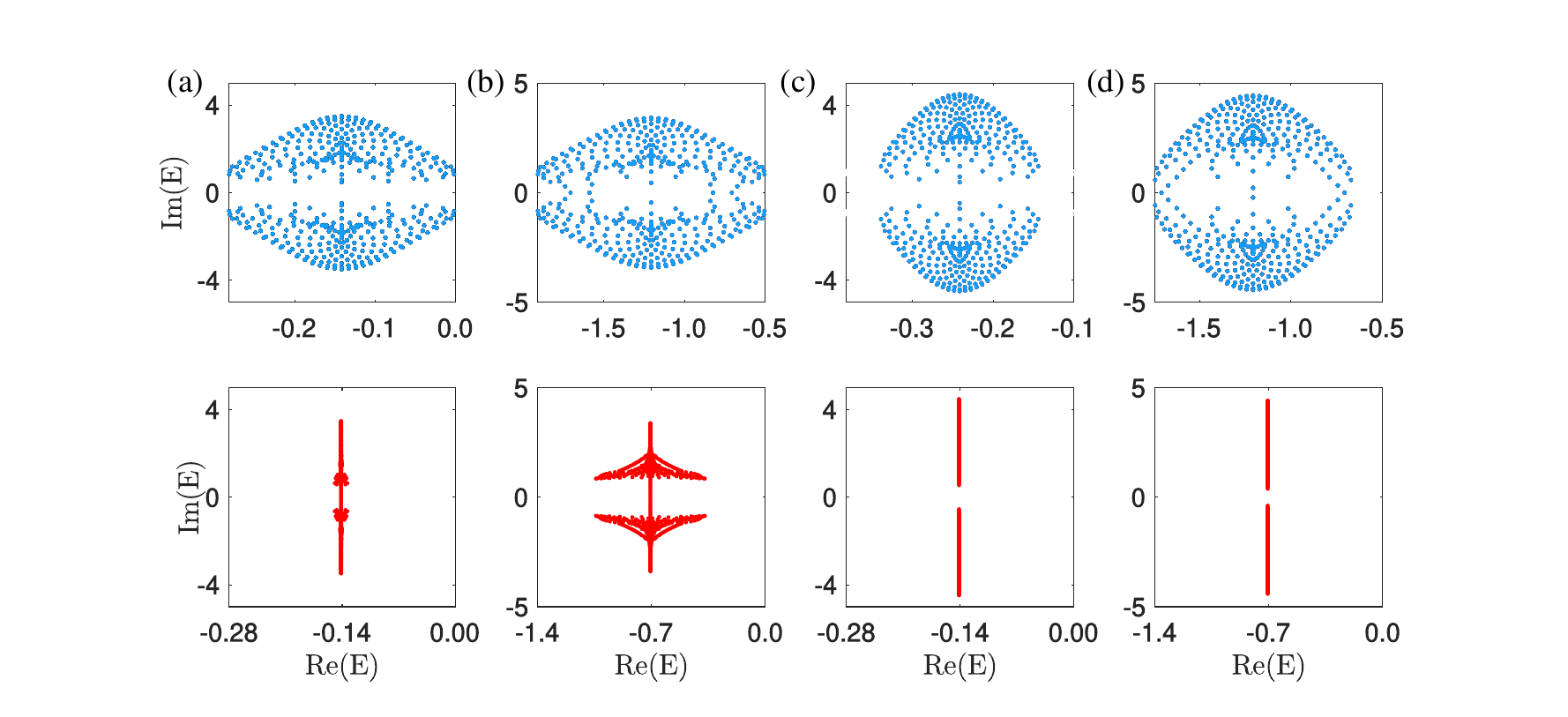}
	\caption{\label{fig01} 
		The eigenvalues of the damping matrix X. Blue: periodic boundary; Red: open boundary . The Liouvillian gap under periodic boundary condition vanishes for (a) and (b) while it is nonzero for (c) and (d). Under open boundary condition, the Liouvillian gap is nonzero in all four cases. This significant difference between open and periodic boundary comes from the NHSE of $X$. (a)$\lambda =0.1,m=1.5$;(b)$\lambda =0.5,m=1.5$; (c)$\lambda =0.1,m=2.5$; (d)$\lambda =0.5,m=2.5$. 
	}
\end{figure*}
\section{General formalism of damping matrix} 
An open quantum system undergoing Markovian damping satisfies the Lindblad master equation
\begin{align}
	\begin{split}
\frac{d\rho }{dt}=-i[H,\rho ]+\sum{(2{{L}_{\mu }}\rho L_{\mu }^{\dagger }-\{L_{\mu }^{\dagger }{{L}_{\mu }},\rho \})},
	\end{split}
\end{align}
where $\rho$ is the density matrix of the system, $H$ is the Hamiltonian that represents unitary evolution of the system, and ${{L}_{\mu }}$ are Lindblad dissipation operators describing the quantum jumps induced by the coupling to the environment. The above equation can be abbreviated as $\frac{d\rho }{dt}=\mathcal{L}\rho$, where $\mathcal{L}$ is called the Liouvillian superoperator. By regarding the density matrix $\rho$ as a vector that consists of matrix elements ${{\rho }_{i,j}}$,$\mathcal{L}$   is represented as a matrix whose elements are given by\cite{okuma2021quantum}
\begin{widetext}
\begin{align}
	\begin{split}
		{{\mathcal{L}}_{ij,kl}}=\sum\limits_{\mu }{2{{L}_{\mu ;i,k}}L_{\mu ;l,j}^{\dagger }-i{{(H-i\sum\limits_{\mu }{L_{\mu }^{\dagger }{{L}_{\mu }}})_{i,k}}}{{\delta }_{l,j}}+i{{(H+i\sum\limits_{\mu }{L_{\mu }^{\dagger }{{L}_{\mu }}})_{l,j}}}{{\delta }_{ik.}}}.
	\end{split}
\end{align}
\end{widetext}
These representations enable one to treat the Lindblad equation as a linear equation. In other words, the dynamics of the system can be understood in terms of the eigenvalue problem of the Liouvillian matrix:$\mathcal{L}{{\rho }^{(i)}}={{\lambda }_{i}}{{\rho }^{(i)}}.$The Hamiltonian and dissipators can be expressed in terms of 2n Majorana fermions\cite{prosen2008third}
\begin{align}
	\begin{split}
		H=\sum\limits_{i,j=1}^{2n}{{{\gamma }_{i}}H_{_{ij}}^{M}{{\gamma }_{j}}},{{L}_{\mu }}=\sum\limits_{i=1}^{2n}{l_{_{\mu ,i}}^{M}{{\gamma }_{i}}}
	\end{split}
\end{align}
where ${{\gamma }_{i}}$ are Majorana fermions satisfying $\{{{\gamma }_{i}},{{\gamma }_{j}}\}=2{{\delta }_{ij}}$. The matrix ${{H}^{M}}$ is chosen to be an antisymmetric matrix,${{\left( {{H}^{M}} \right)}^{T}}=-{{H}^{M}}$. Defining ${{M}_{ij}}=\sum\nolimits_{\mu }{l_{\mu ,i}^{*}l_{\mu ,j}^{{}}},M_{_{ij}}^{M}=\sum\nolimits_{\mu }{{{(l_{\mu ,i}^{M})}^{*}}l_{\mu ,j}^{M}}$, we have $M_{_{{}}}^{M}=\frac{1}{4}M\otimes (1+{{\sigma }_{y}})$. Under the third quantization\cite{lieu2020tenfold,prosen2008third}, the Liouvillian superoperator is expressed as a quadratic form of the 2n complex fermions (4n Majorana fermions) 
\begin{align}
	\begin{split}
		\begin{aligned}
		\mathcal{L}\text{=}\frac{2}{i}\left( \begin{matrix}
			{{c}^{\dagger }} & c  \\
		\end{matrix} \right)\left( \begin{matrix}
			-{{Z}^{T}} & Y  \\
			0 & Z  \\
		\end{matrix} \right)\left( \begin{matrix}
			c  \\
			{{c}^{\dagger }}  \\
		\end{matrix} \right),	
		\end{aligned}	
	\end{split}
\end{align}
where $Z={{H}^{M}}+i\operatorname{Re}{{({{M}^{M}})}^{T}},Y=2\operatorname{Im}{{({{M}^{M}})}^{T}},$and $c=({{c}_{1}},{{c}_{2}},...,{{c}_{2n}})$ are third quantized complex fermions. Through the above expression,we can obtain the Liouvillian eigenspectrum\cite{lieu2020tenfold,prosen2008third}
\begin{align}
	\begin{split}
		\lambda =\sum\nolimits_{i}{{{E}_{i}}{{v}_{i}}} 
	\end{split}
\end{align}
with ${{v}_{i}}\in \{0,1\}$,where $\{{{E}_{i}}\}$ is the eigenspectrum of $4iZ$.Here $\lambda$ contains valuable information of the full density-matrix dynamics, and it can be easily obtained from the damping matrix X with  ${{X}_{ij}}=i{{h}_{ji}}-\sum\nolimits_{\mu }{l_{\mu ,j}^{*}l_{\mu ,i}^{{}}}$\cite{li2022engineering}.Rewriting M as $M={{M}_{r}}+i{{M}_{i}}$,where ${{M}_{r}},{{M}_{i}}$ are real matrices,we have ${{M}^{M}}=\frac{1}{4}({{M}_{r}}+i{{M}_{i}})\otimes (1+{{\sigma }_{y}})$.  $Z$ can be further written as
\begin{align}
	\begin{split}
			 Z&=\frac{1}{4}({{h}_{r}}\otimes {{\sigma }_{y}}+i{{h}_{i}}\otimes 1)+i\frac{1}{4}(M_{r}^{T}\otimes 1-iM_{i}^{T}\otimes {{\sigma }_{y}}) \\ 
			& =\frac{1}{4}({{h}_{r}}+M_{i}^{T})\otimes {{\sigma }_{y}}+i\frac{1}{4}({{h}_{i}}+M_{r}^{T})\otimes 1  
	\end{split}
\end{align}
Therefore,
\begin{widetext}
\begin{align}
	\begin{split}
		& \det (4iZ-\lambda E)=\det \left( \begin{matrix}
			-({{h}_{i}}+M_{r}^{T})-\lambda  & {{h}_{r}}+M_{i}^{T}  \\
			-({{h}_{r}}+M_{i}^{T}) & -({{h}_{i}}+M_{r}^{T})-\lambda   \\
		\end{matrix} \right) =\det (X-\lambda E)\det ({{X}^{*}}-\lambda E)
	\end{split}
\end{align}
\end{widetext}
So,the eigenvalue of $4iZ$ are the union of the eigenvalues of $X$ and ${{X}^{*}}$,which gives the Liouvillian eigenspectrum.

Then we outline the general form of the Lindblad damping matrix in open quantum systems\cite{song2019non}.We consider tight-binding models whose Hamiltonian can generally be written as $H=\sum{_{ij}{{h}_{ij}}c_{i}^{\dagger }}{{c}_{j}}$,where $c_{i}^{\dagger },{{c}_{i}}$ are the creation and annihilation operators on lattice site $i$, and ${{h}_{ij}}=h_{ij}^{*}$ is the hopping amplitude between the lattice points of the system ($i\ne j$) or onsite potential ($i=j$). It is convenient to define the single-particle correlation function ${{\Delta }_{ij}}(t)=Tr[c_{i}^{\dagger }{{c}_{j}}\rho (t)]$ to observe the time evolution of the density matrix. Each cell is coupled to the environment through the gain jump operator $L_{\mu }^{g}=\sum{_{i}D_{\mu i}^{g}}c_{i}^{\dagger }$ and loss jump operator $L_{\mu }^{l}=\sum{_{i}D_{\mu j}^{l}}{{c}_{i}}$. Substituting the Lindblad quantum master equation into the time evolution of the single-particle correlation function, we can obtained
\begin{align}
	\begin{split}
		\frac{d\Delta (\text{t})}{dt}=X\Delta (\text{t})+\Delta (\text{t}){{X}^{\dagger }}+2{{M}_{g}},
	\end{split}
	\end{align}
	where $X=i{{h}^{T}}-(M_{l}^{T}+{{M}_{g}})$ is the damping matrix with ${{({{M}_{g}})}_{i}}_{j}=\sum{_{\mu }D_{\mu i}^{g*}}D_{\mu j}^{g}$ and ${{({{M}_{l}})}_{i}}_{j}=\sum{_{\mu }D_{\mu i}^{l*}}D_{\mu j}^{l}$.The steady state correlation ${{\Delta }_{s}}=\Delta (\infty )$, to which the long-time evolution of any initial state converges, is determined by $d{{\Delta }_{s}}/dt=0$ or $X{{\Delta }_{s}}+{{\Delta }_{s}}{{X}^{\dagger }}+2{{M}_{g}}=0$. Focusing on the deviation towards the steady state $\tilde{\Delta }(t)=\Delta (t)-{{\Delta }_{s}}$ whose time evolution is ${d\tilde{\Delta }(t)}/{dt}\;=X\tilde{\Delta }(t)+\tilde{\Delta }(t){{X}^{\dagger }}$.we can integrate it with Eq. (1) to obtain
	\begin{align}
		\begin{split}
			\tilde{\Delta }(\text{t})={{e}^{Xt}}\tilde{\Delta }(\text{0}){{e}^{X\dagger t}}.
		\end{split}
	\end{align}
	Therefore, the dynamical behaviour of the system can be characterized by the damping matrix.

\begin{figure}[]
	\includegraphics[width=0.45\textwidth]{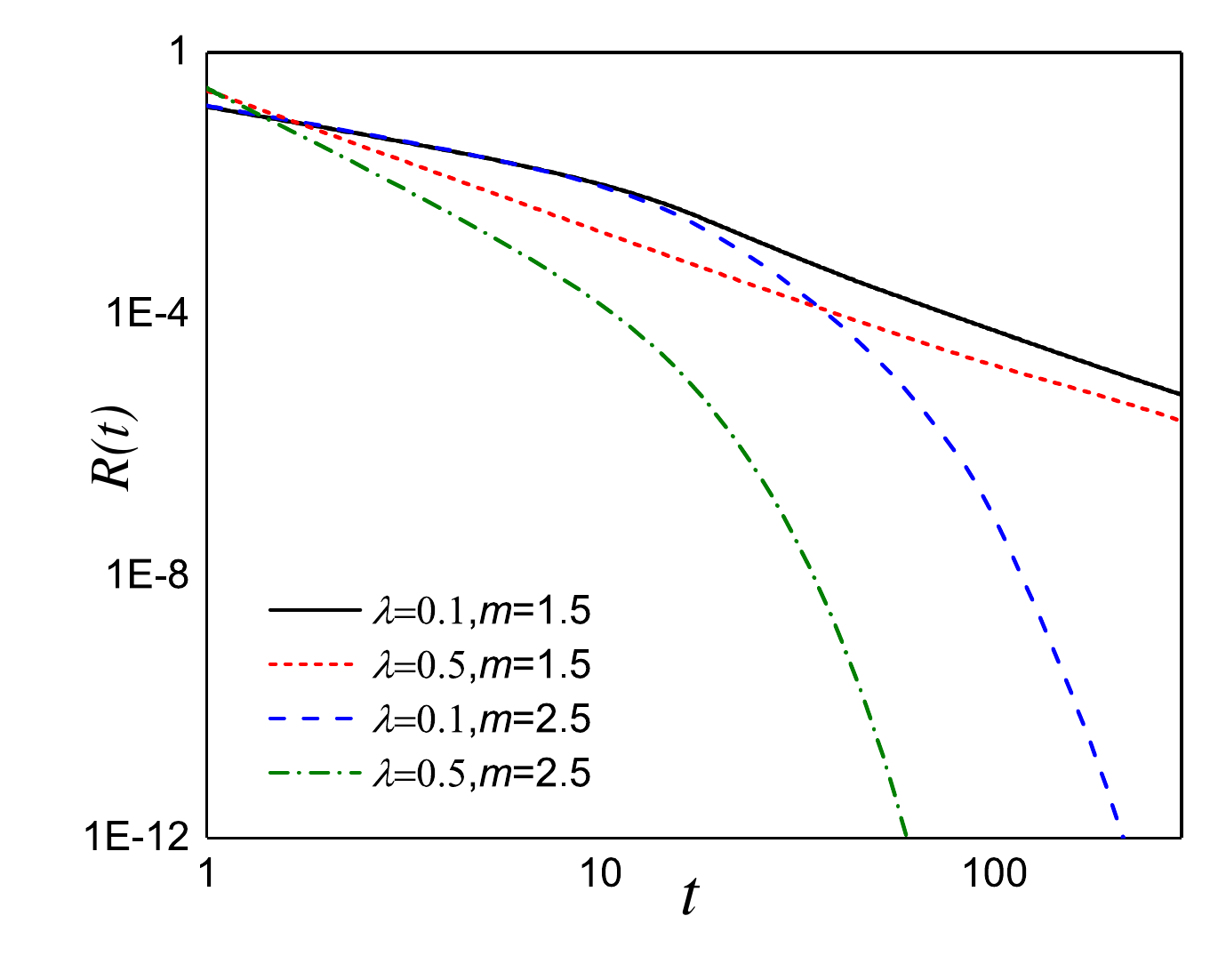}
	\caption{\label{fig01} 
		The damping of site-averaged fermion number towards the steady state under periodic boundary condition with size $L=30\times 30$. $m = 1.5$ (black and red) exhibits a slow algebraic damping while $m = 2.5$ (blue and green) is an exponential damping. The  initial state is the completely filled state.
	}
\end{figure}
\begin{figure*}[]
	\includegraphics[width=0.98\textwidth]{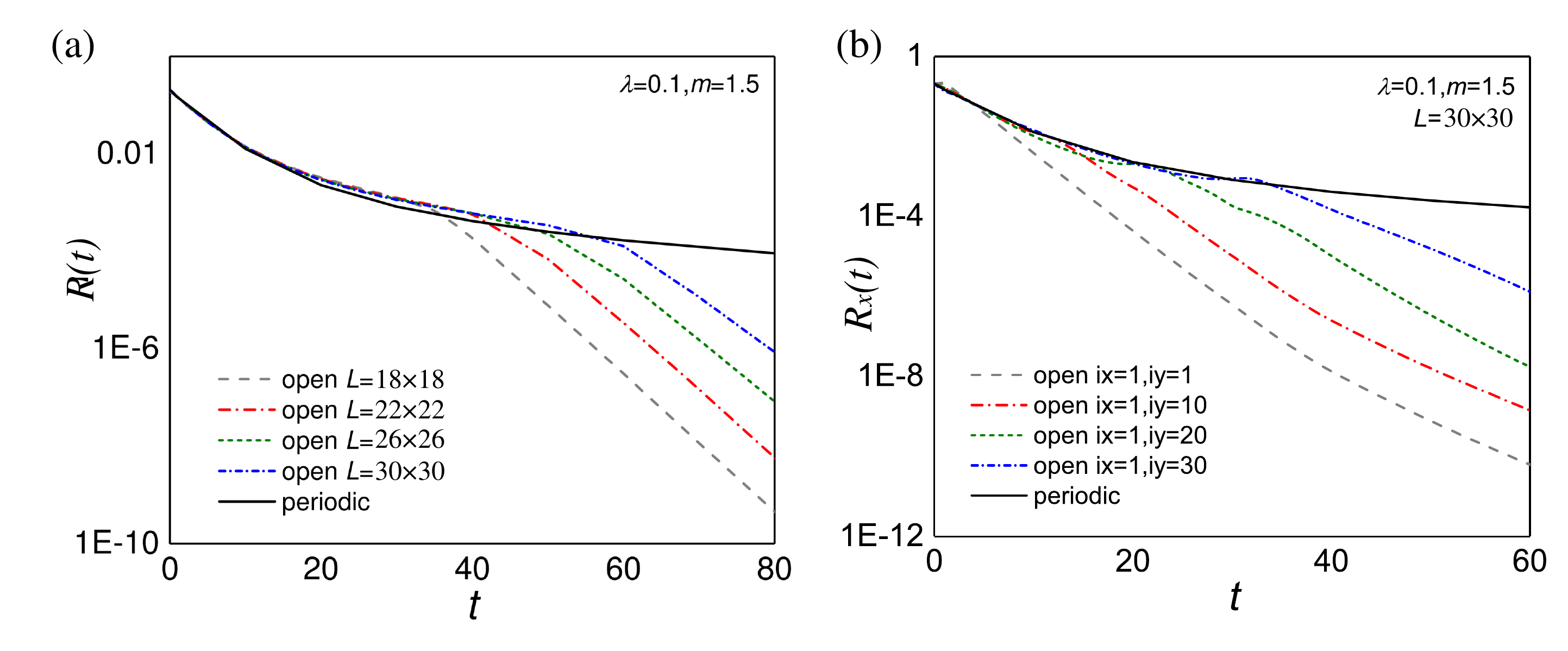}
	\caption{\label{fig03} 
		(a) The site-averaged particle number damping under periodic boundary conditions (solid line) and open boundary conditions (dashed line) for several sizes $L$. The long-time damping of $R(t)$ follows a power law under periodic boundary condition, while the damping follows an exponential law after an initial power law stage under open boundary condition. (b) The particle number damping on several sites. The system size is 30×30, and the left end ($ix = 1, iy = 1$) enters the exponential phase from the beginning, followed by the other sites in turn. For (a) and (b), the initial state is completely filled state, $m=1.5,\lambda =0.1$.
	}
\end{figure*}
\section{Model} 
In this paper, we consider the Chen insulator model with the Hamiltonian in momentum space as
\begin{align}
	\begin{split}
		h(k)={{l}_{x}}\sin {{k}_{x}}{{\sigma }_{x}}+{{l}_{y}}\sin {{k}_{y}}{{\sigma }_{y}}+{{\varepsilon }_{k}}{{\sigma }_{z}},
	\end{split}
\end{align}
where ${{\varepsilon }_{k}}=m+{{t}_{x}}\cos {{k}_{x}}+{{t}_{y}}\cos {{k}_{y}}$.Let each unit cell contain a single loss and gain dissipator,
\begin{align}
	\begin{split}
		\begin{aligned}
			& L_{x}^{l}=\frac{\sqrt{2\gamma }}{2}({{e}^{-\frac{\pi }{4}i}}{{c}_{xA}}+{{e}^{\frac{\pi }{4}i}}{{c}_{xB}}) \\ 
			& L_{x}^{g}=\frac{\sqrt{2\gamma }}{2}({{e}^{\frac{\pi }{4}i}}c_{xA}^{\dagger }+{{e}^{-\frac{\pi }{4}i}}{{c}_{xB}}), \\ 
		\end{aligned}		
	\end{split}
\end{align}
where $x$ denotes the lattice site, $A,B$ refer to the sublattice. The Fourier transformation of $X$ is $X(k)=i{{h}^{T}}(-k)-M_{l}^{T}(-k)-{{M}_{g}}(k)$. The gain and loss dissipators are intra-cell, so these $M(k)$ matrices are independent of $k$, ${{M}_{l}}(k)=\frac{\sqrt{2}}{2}\lambda +\frac{1}{2}{{\sigma }_{x}}-\frac{1}{2}{{\sigma }_{y}},{{M}_{g}}(k)=\frac{\sqrt{2}}{2}\lambda +\frac{1}{2}{{\sigma }_{x}}+\frac{1}{2}{{\sigma }_{y}}$. Then, the damping matrix in momentum space is
	\begin{align}
	\begin{split}
		X(k)=i[{{l}_{x}}\sin {{k}_{x}}{{\sigma }_{x}}+{{l}_{y}}\sin {{k}_{y}}{{\sigma }_{y}}+{{\varepsilon }_{k}}{{\sigma }_{z}}+i[\lambda {{\sigma }_{x}}+\lambda {{\sigma }_{y}}]-\sqrt{2}\lambda.
	\end{split}
\end{align}
It can be written in the form of left and right eigenvectors,
	\begin{align}
	\begin{split}
		X=\sum\limits_{n}{{{\lambda }_{n}}|{{u}_{Rn}}\rangle \langle {{u}_{Ln}}|},
	\end{split}
\end{align}
where ${{X}^{\dagger }}|{{u}_{Ln}}\rangle =\lambda _{n}^{*}|{{u}_{Ln}}\rangle, X|{{u}_{Rn}}\rangle ={{\lambda }_{n}}|{{u}_{Rn}}\rangle$. It is worth noting that our ${{M}_{l}}$ andr ${{M}_{g}}$ satisfy $M_{l}^{T}+{{M}_{g}}=2{{M}_{g}}$, guaranteeing that ${{\Delta }_{S}}=\frac{1}{2}{{I}_{2L\times 2L}}$ is a steady state solution , where $L={{N}_{x}}\times {{N}_{y}}$, $L$ is the system size, and ${{N}_{x}},{{N}_{y}}$ are the size in $x,y$ direction respectively. We assume that the initial state of the system is the completely filled state, i.e.,$\Delta (0)$ is an identity matrix. Therefore, Eq.(4) can be re-expressed as
\begin{widetext}
\begin{align}
	\begin{split}
		\begin{aligned}
			 \tilde{\Delta }(\text{t})&=\frac{1}{2}\sum\limits_{n,{n}',l}{\exp [({{\lambda }_{n}}+\lambda _{{{n}'}}^{*})t]\widetilde{{{u}_{R}}}(i,n)}\widetilde{{{u}_{L}}}(l,n)\widetilde{u_{L}^{*}}(l,{{n}^{'}})\widetilde{u_{R}^{*}}(j,{{n}^{'}}) \\ 
			& =\frac{1}{2}\sum\limits_{n,{{n}^{'}}}{\frac{\sum\nolimits_{l}{\exp [({{\lambda }_{n}}+\lambda _{{{n}'}}^{*})t]{{u}_{R}}(i,n){{u}_{L}}(l,n)u_{L}^{*}(l,{{n}^{'}})u_{R}^{*}(j,{{n}^{'}})}}{\sum\nolimits_{k}{{{u}_{R}}(k,n){{u}_{L}}(k,n)\sum\nolimits_{m}{u_{L}^{*}(m,{{n}^{'}})u_{R}^{*}(m,{{n}^{'}})}}}}  
		\end{aligned}		
	\end{split}
\end{align}
\end{widetext}
According to the dissipative property, $\operatorname{Re}\{{{\lambda }_{n}}\}\le 0$ always holds. The Liouvillian gap $\Lambda =\min [2\operatorname{Re}(-{{\Lambda }_{n}})]$ plays a decisive role in long-time dynamics. The opening gap ($\Lambda \ne 0$) implies an exponential rate of convergence to the steady state, while the closing gap($\Lambda =0$) implies algebraic convergence\cite{cai2013algebraic}.
\begin{figure}[]
	\includegraphics[width=0.45\textwidth]{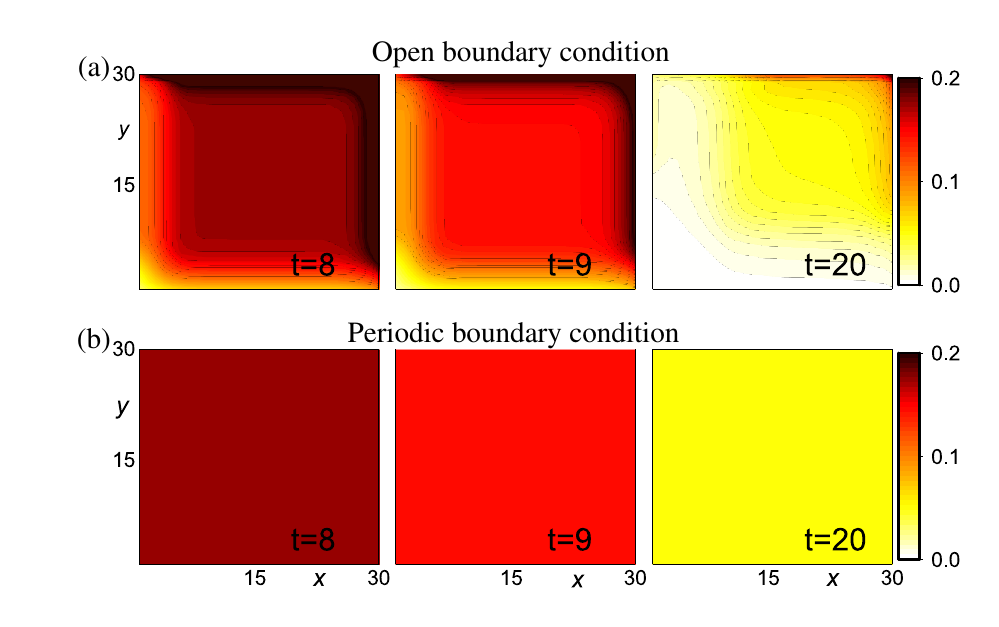}
	\caption{\label{fig03} 
		The evolution of $R(t)$ at each lattice site under open boundary conditions (a) and periodic boundary conditions (b).
	}
\end{figure}

\section{Chiral damping} 
For simplicity, the parameters of our model are taken as ${{l}_{x}}={{l}_{y}}=1,{{t}_{x}}={{t}_{y}}=-1.$. We first study the dynamical behaviour under the periodic boundary conditions. Diagonalizing $X(k)$, we obtain the energy spectrum as shown in Fig. 1. It is found that the Liouvillian gap vanishes at $m=1.5$, while the gap opens at $m=2.5$. So we expect the damping rate to be algebraic and exponential in each case, respectively.

To verify this, we define the site-averaged fermion number deviation from the steady state $R(t)=\sqrt{\frac{1}{{{N}_{X}}{{N}_{Y}}}{{\sum\limits_{x}{{{R}_{x}}(t)}}^{2}}}=\sqrt{\frac{1}{{{N}_{X}}{{N}_{Y}}}{{\sum\limits_{x}{\left( \frac{{{n}_{x}}(t)-{{n}_{x}}(t-{{\delta }_{t}})}{{{\delta }_{t}}} \right)}}^{2}}}$, where ${{R}_{x}}(t)={{n}_{x}}(t)-{{n}_{x}}(\infty )$, ${{n}_{x}}(t)={{\Delta }_{xA,xA}}(t)+{{\Delta }_{xB,xB}}(t)$. The numerical results are shown in Fig. 2. As anticipated, it is observed that the damping of $R(t)$ is algebraic for cases black and red lines with $m=1.5$ while exponential for blue and green lines with $m=2.5$ under the periodic boundary condition.

Next we turn to the open boundary conditions. Since the damping matrix $X$ has  NHSE, its energy spectrum is no longer that of the periodic boundary conditions. At this point the energy spectrum has a non-zero energy gap (blue part of Fig. 1) ,therefore, we expect an exponential long-time damping of $\tilde{\Delta }(t)$. The numerical simulation in Fig. 3 confirms this exponential behaviour with $R(t)$ having a period of algebraic damping before entering into the exponential damping. The time of the algebraic damping increases with the size $L$[Fig.3(a)]. To better understand this feature, we plot the damping in several unit cells in the same $x$ dimension ($ix = 1$), as shown in Fig. 3(b). It can be seen that the left end ($iy = 1$) enters the exponential damping immediately, and the other sites enter the exponential damping in turn according to their different distances to the left end. Due to a process of algebraic damping that occurs before entering the exponential stage, there is a "damping wavefront" from left ($ix = 1, iy = 1$) to right ($ix=1,iy={{N}_{y}}$). This phenomenon is known as "chiral damping".

The phenomenon of chiral damping can be observed more intuitively as shown in Fig. 4(a) where the colour shades indicate the value of $R(t)$. Under the periodic boundary condition, the time evolution follows a slow power law while under the open boundary condition, a wavefront moving to the upper right is observed. This can be intuitively linked to the phenomenon that all eigenstates of $X$ are localized in the upper right corner, which arises from the non-Hermitian skin effect of the damping matrix $X$. If the matrix $X$ does not have NHSE under the open boundary condition, the fermion number of the system should have a similar behaviour of damping under different boundary conditions. Therefore, the non-Hermitian skin effect plays an important role in open quantum systems and significantly affects the dynamical behaviour of open quantum systems.

\section{Experiment realized} 
Next we give the scheme of topolectrical circuits to simulate the damping matrix. Based on the similarity between the Kirchhoff equation and the Schrödinger equation, it is possible to simulate the Hamiltonian of the system using different circuit components, and the different parameters in the Hamiltonian can be adjusted independently by various components. The circuit Laplacian corresponding to the Hamiltonian can be written as
	\begin{align}
	\begin{split}
		J=D-C+W,
			\end{split}
	\end{align}
	where $W$ and $D$ are diagonal matrices containing the total conductance from each node to the ground and to the rest of the circuit, respectively. $C$ is the adjacency matrix of conductances\cite{lee2018topolectrical}.

	Considering the periodic boundary conditions first, the topolectrical circuit for realizing the damping matrix X is illustrated in Fig.5.  Fig. 5(a) depicts the schematic diagram of the overall circuit structure, which gives the connection relationship between the nodes. Fig. 5(b) shows the detailed circuit component of the unit which is the green dashed box in Fig. 5(a). The blue box in Fig. 5(a) represents a unit cell in the system, and the two nodes inside it correspond to sublattices $A$ (red) and $B$ (blue). The circuit connections in the x and y directions are distinguished by black and gray. From Fig. 5(b) we can obtain the matrices $C$ and $D$ in Eq. (9), so that
\begin{figure*}[]
	\includegraphics[width=0.98\textwidth]{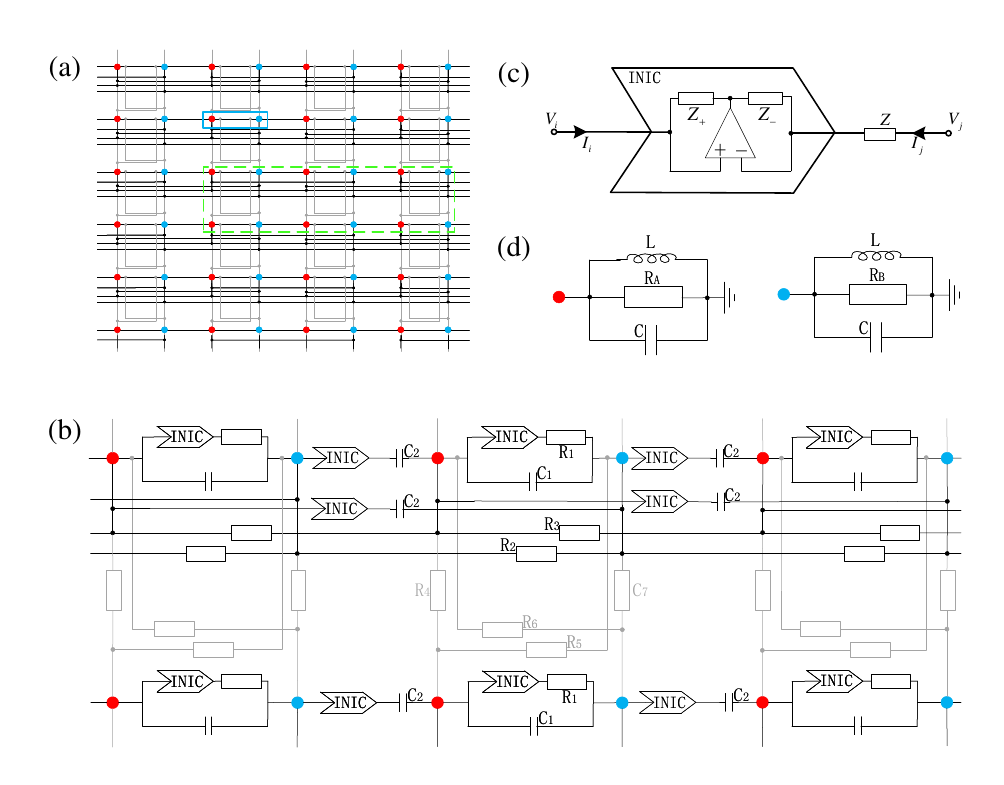}
	\caption{\label{fig05} 
		The structure of topolectrical circuit to realize damping matrix under periodic boundary conditions.  (a) Connection relations between the nodes. The blue solid line box containing two “sublattice” nodes $A$ (red) and $B$ (blue) simulates a unit cell of $X$. The black(grey) solid line indicates the coupling between nodes in the $x$($y$)-direction. (b)The circuit element structure is detailed for the green dashed framed rectangle in (a).    (c) Internal circuit diagram of the INIC element, consisting of an operational amplifier and impedances ${{Z}_{\pm }}$. The impedance $Z$ is the target element, and different conductance in different directions of ${{V}_{i,r}}$ can be achieved by connecting the INIC in series. ${{Z}_{\pm }}$ satisfies ${{Z}_{+}}={{Z}_{-}}$. (d) Grounding module of the nodes. The resistances ${{R}_{A,}}{{R}_{B}}$ and capacitance $C$ are used to simulate the onsite potential, and inductance $L$ allows the Laplacian eigenvalue spectrum to be shifted uniformly as desired.
	}
\end{figure*}

\begin{widetext}
	\begin{align}
		\begin{split}
	\begin{aligned}
		& D-C=-i\omega \left( \begin{matrix}
			-i\frac{2}{\omega {{R}_{2}}}\cos {{k}_{x}}-i\frac{2}{\omega {{R}_{4}}}\cos {{k}_{y}}-{{C}_{1}}-i\frac{1}{\omega {{R}_{0}}} & i2{{C}_{2}}\sin {{k}_{x}}+\frac{2}{\omega {{R}_{6}}}\sin {{k}_{y}}-i\frac{1}{\omega {{R}_{1}}}+{{C}_{1}}  \\
			i2{{C}_{2}}\sin {{k}_{x}}+\frac{2}{\omega {{R}_{5}}}\sin {{k}_{y}}+i\frac{1}{\omega {{R}_{1}}}+{{C}_{1}} & -i\frac{2}{\omega {{R}_{3}}}\cos {{k}_{x}}-i\frac{2}{\omega {{R}_{7}}}\cos {{k}_{y}}-{{C}_{1}}+i\frac{1}{\omega {{R}_{0}}}  \\
		\end{matrix} \right), \\ 
		& \text{    } \\ 
	\end{aligned}
		\end{split}
\end{align}
\end{widetext}
	with ${{C}_{1}}=-\lambda ,{{C}_{2}}={{{l}_{x}}}/{2}\;,\frac{1}{{{R}_{0}}}=\frac{1}{{{R}_{1}}}-2(\frac{1}{{{R}_{3}}}+\frac{1}{{{R}_{4}}}),{{R}_{1}}=-{1}/{(\omega \lambda )}\;,{{R}_{2}}=-{{R}_{3}}=-{2}/{{{t}_{x}}}\;$, ${{R}_{4}}=-{{R}_{7}}=-{2}/{{{t}_{y}}}\;$, ${{R}_{5}}=-{{R}_{6}}=-{2}/{{{l}_{y}}}\;.$
	
	Comparing it with the damping matrix, we need to add grounding elements to match the onsite potential. The grounding elements of nodes $A$ and $B$ are shown in  Fig. 5(d), where the resistors ${{R}_{A,}}{{R}_{B}}$ and capacitors $C$ simulate the lattice potential , and ${{R}_{A,}}{{R}_{B}}$ satisfies ${{R}_{A}}=-{{R}_{B}}$. So the diagonal matrix $W$ is
		\begin{align}
		\begin{split}
			W=\left( \begin{matrix}
				i\omega C+\frac{1}{{{R}_{A}}}+\frac{1}{i\omega L} & {}  \\
				{} & i\omega C+\frac{1}{{{R}_{B}}}+\frac{1}{i\omega L}  \\
			\end{matrix} \right).
		\end{split}
	\end{align}
	From Eq. (10) we get the conductance matrix of the circuit of Fig. 5(a) at $\omega$ frequency
\begin{widetext}
		\begin{align}
		\begin{split}
	\begin{aligned}
		 J(\omega )&=-i\omega \left( \begin{matrix}
			-i\frac{2}{\omega {{R}_{2}}}\cos {{k}_{x}}-i\frac{2}{\omega {{R}_{4}}}\cos {{k}_{y}}-(C+{{C}_{1}})-i(\frac{1}{\omega {{R}_{0}}}+\frac{1}{\omega {{R}_{A}}}) & i2{{C}_{2}}\sin {{k}_{x}}+\frac{2}{\omega {{R}_{6}}}\sin {{k}_{y}}-i\frac{1}{\omega {{R}_{1}}}+{{C}_{1}}  \\
			i2{{C}_{2}}\sin {{k}_{x}}+\frac{2}{\omega {{R}_{5}}}\sin {{k}_{y}}+i\frac{1}{\omega {{R}_{1}}}+{{C}_{1}} & -i\frac{2}{\omega {{R}_{3}}}\cos {{k}_{x}}-i\frac{2}{\omega {{R}_{7}}}\cos {{k}_{y}}-(C+{{C}_{1}})+i(\frac{1}{\omega {{R}_{0}}}+\frac{1}{\omega {{R}_{A}}})  \\
		\end{matrix} \right)\\&+\frac{1}{i\omega L}\varepsilon  \\ 
		& =-i\omega {{J}_{P}}+\frac{1}{i\omega L}\varepsilon ,  
	\end{aligned}
			\end{split}
\end{align}
\end{widetext}
\begin{figure*}[]
	\includegraphics[width=0.98\textwidth]{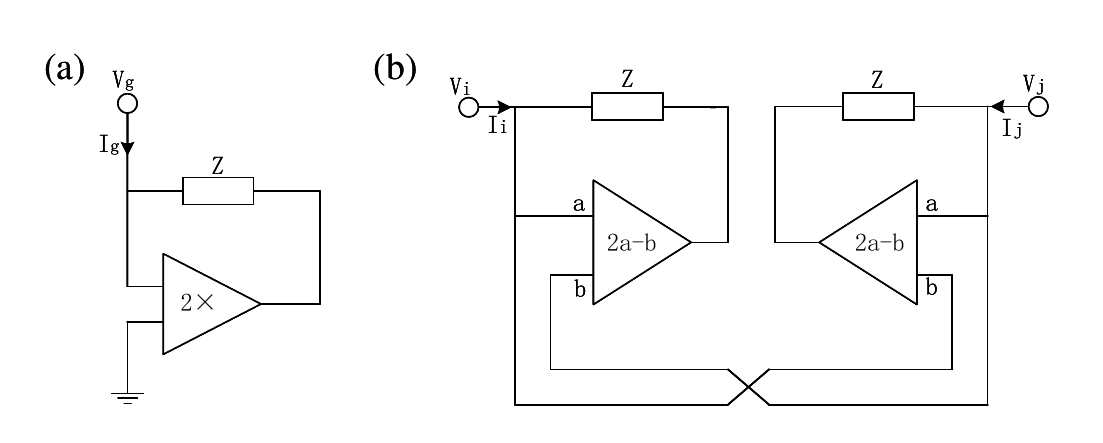}
	\caption{\label{fig06} 
		Negative impedance module\cite{wang2018topological}. (a) A single-port circuit to ground. The input impedance is ${{Z}_{g}}=-Z$; (b) Free-port circuit. Its input impedance at both ends is ${{Z}_{ij}}={{Z}_{ji}}=-Z$. The markings on the ideal amplifier indicate the output voltage versus the input voltage.
	}
\end{figure*}
\begin{figure*}[ ]
	\includegraphics[width=0.98\textwidth]{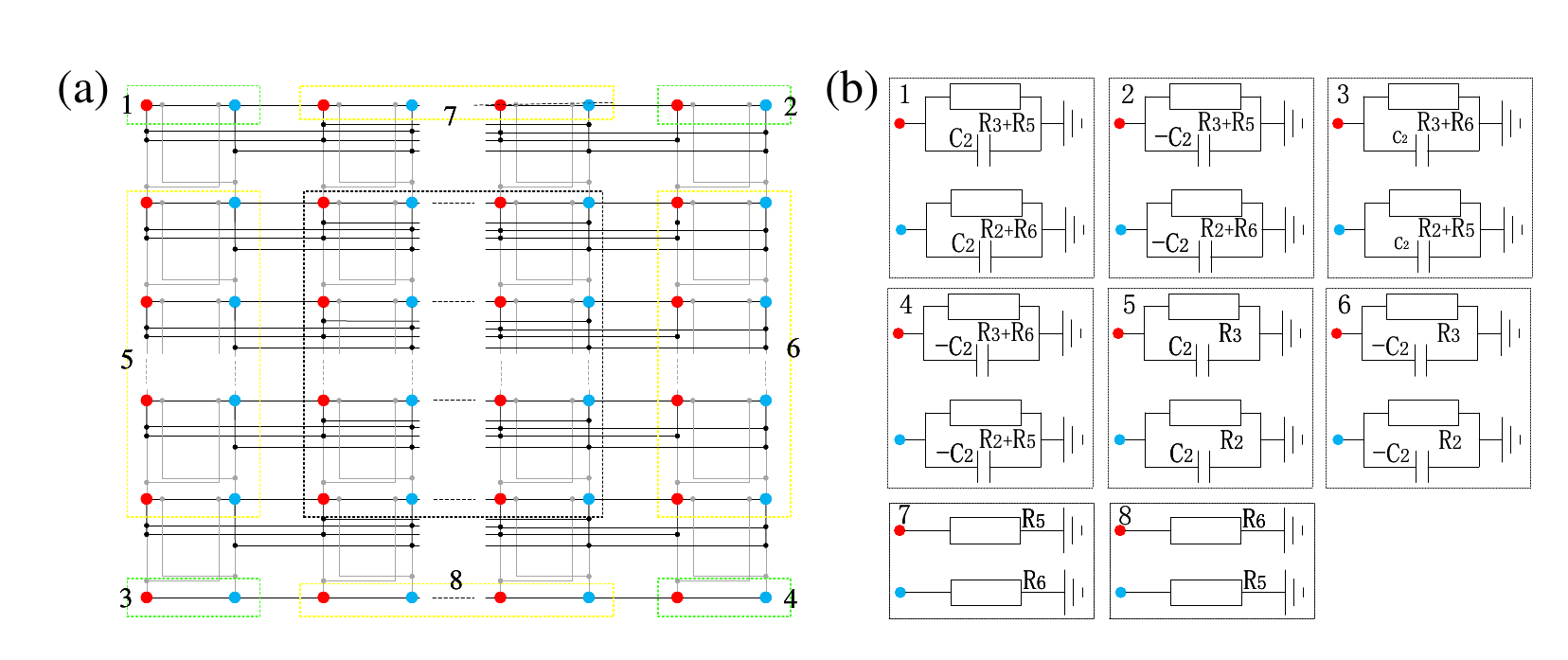}
	\caption{\label{fig01} 
		Schematic diagram of the circuit of the damping matrix X under open boundary conditions. (a) Schematic diagram of the connection relations among the nodes. The black, yellow and green dashed boxes correspond to the body, edge and corner nodes, respectively. The circuit connections of the body node are the same as those of the periodic boundary, while the edge and corner nodes require additional grounding elements to regulate the onsite potential. (b) Grounding modules for edge and corner nodes . The grounding elements for the edge and corner nodes are different for each of the four orientations, where the negative impedance elements can be realized by Fig.6(a). Note that in addition to these grounding elements, all nodes need to be connected to the elements in Fig. 5(d).}
\end{figure*}
where $C=(1-\sqrt{2})\lambda$, $\frac{1}{{{R}_{A}}}=-\frac{1}{{{R}_{B}}}=-\omega m-\frac{1}{{{R}_{0}}}.$ Comparing this Laplacian matrix with the damping matrix, the mapping relationship can be established by ${{J}_{P}}\Leftrightarrow X$.
	
Notice that the circuit requires a negative component, which is implemented as shown in Fig.6. Fig.6(a) and (b) show the equivalent negative impedance modules for a single port to ground and a free two-terminal port, respectively. They achieve the equivalent negative impedance through an amplifier. According to Kirchhoff's law, the input impedance of the single-port circuit to ground [Fig. 6(a)] can be obtained as 
\begin{align}
	\begin{split}
{{Z}_{g}}=\frac{{{V}_{g}}}{{{I}_{g}}}=\frac{{{V}_{g}}}{({{V}_{g}}-2{{V}_{g}})/Z}=-Z.
\end{split}
\end{align}
The input impedance at both ends of the free port circuit [Fig. 6(b)] are
\begin{align}
	\begin{split}
\begin{aligned}
	& {{Z}_{ij}}=\frac{{{V}_{i}}-{{V}_{j}}}{{{I}_{i}}}=\frac{{{V}_{i}}-{{V}_{j}}}{[{{V}_{i}}-2({{V}_{i}}-{{V}_{j}})]/Z}=-Z, \\ 
	& {{Z}_{ji}}=\frac{{{V}_{j}}-{{V}_{i}}}{{{I}_{j}}}=\frac{{{V}_{j}}-{{V}_{i}}}{[{{V}_{j}}-2({{V}_{j}}-{{V}_{i}})]/Z}=-Z. \\ 
\end{aligned}
\end{split}
\end{align}
That is ${{Z}_{ij}}={{Z}_{ji}}=-Z$.

Under the open boundary condition, the hopping amplitude of the cells located at the boundary weakens, leading to fewer branches connected to the boundary nodes in the circuit model, as shown in Fig.7(a). Fig.7(a) gives the connection relationship between the nodes of the circuit under the open boundary condition, and the circuit nodes can be classified into body nodes (in the black dashed box), edge nodes (in yellow) and corner nodes (in green). Changes in the branch circuit of the nodes at the boundary will cause variations of the matrices $C$ and $D$. The matrix $C$ corresponds to the hopping amplitude between the lattice points, which is allowed to change. Whereas the change of $D$ is not desired due to the same onsite potential under different boundary condition. 

Therefore, we need to design specific grounding elements to eliminate the effects of variations in $D$. Owing to the asymmetry of the coupling strengths under periodic boundary condition, the types of the edge and corner nodes are different for each of the four orientations, so there are a total of 16 different grounding modules, as shown in Fig. 7(b). The additional grounding elements keep the diagonal matrix $D+W$ unchanged, i.e., the onsite potential is unchanged, which achieves the mapping of the circuit Laplacian in Fig. 7 to the damping matrix $X$ under the open boundary condition.
\section{Conclusion} 
In summary, we study the dynamical properties of a two-dimensional open system. The open quantum system is constructed by introducing appropriate gain and loss to the Chen insulator, and then using the damping matrix derived from the Liouvillian superoperator explore its long-time evolution. It is found that  the site-averaged fermion number deviation from the steady state under periodic boundary conditions shows a slow algebraic damping when the energy gap closes and an exponential damping when the energy gap opens. Under open boundary conditions, due to the non-Hermitian skin effect of the damping matrix, the system exhibits the phenomenon of chiral damping that the fermion number at each site undergoes a period of algebraic damping before entering an exponential damping, and the transition time that is proportional to the distance from that site to the boundary. Finally, we map the damping matrix in terms of the circuit Laplacian to give a model diagram of the topolectrical circuit implementation of the system.

\bibliographystyle{IEEEtran}
\bibliography{manuscript.bbl}

\end{document}